**Paper Number:**
**19-00169**

**Title:**
**CAPTURE AND RECOVERY OF CONNECTED VEHICLE DATA: A COMPRESSIVE
SENSING APPROACH**

**Authors:**
**Lei Lin**
NEXTRANS Center
Purdue University
West Lafayette, IN 47906
Email: lin954@purdue.edu

**Weizi Li**

Department of Computer Science
University of North Carolina at Chapel Hill
201 S Columbia St, Chapel Hill, NC 27599
Email: weizili@cs.unc.edu

**Srinivas Peeta, Corresponding Author**
Lyles School of Civil Engineering
Purdue University
West Lafayette, IN 47906
Email: petta@purdue.edu

## INTRODUCTION

Recent advancements in technology and its implication on socio-economic benefits have prompted widespread research and development of connected vehicles (CVs). These machineries promise to improve the safety and mobility of our transportation system with enhanced situational awareness through vehicle-to-vehicle (V2V) and vehicle-to-infrastructure (V2I) communications.

As an example of such endeavors, the Safety Pilot Model Deployment (SPMD) *(1)* had included nearly 3000 vehicles broadcasting Safety Message (BSM) to nearby vehicles at a rate of 10 Hz. While these CV data bring opportunities for improving intelligent transportation system applications such as traffic state estimation *(2, 3)* and traffic signal optimization *(4)*, the high-sampling rate would result in a large amount of data being captured and stored, leading to prohibitive cost *(5,6)*.

Previous studies have taken two approaches to address this issue. The first approach is called sample-then-compression, which collects data at a fixed rate in real-time and compresses the data offline *(7,8)*. The limitation of this approach is that no optimal solution is provided to adjust the online data sampling rate, hence higher information is still captured, transmitted, and stored. The second approach is to adopt a dynamic perspective by reducing the amount of data captured online while not compromising the system awareness and control requirements of transportation authorities *(9,10)*. The main issue of this approach is that the data collected are limited to specific tasks and time periods, which poses higher requirements for the scalability and stability of data analysis algorithms.

We propose a Compressive-Sensing (CS) based approach for CV data capture and recovery. CS is a novel approach for capturing and recovering signals *(11,12,15,16,17,21)*. Differing from the first approach in which huge amount of data are acquired and compressed, CS enables redundancy removal during the sampling process via a lower but more effective sampling rate *(13)*. Unlike the second approach, CS does not require dynamic adjustment to the data capture rate based on various performance measurement requests *(9)* or traffic control applications *(10)*. Instead, it performs a linear transformation using a Sensing matrix and a Sparsifying matrix to capture the essence of a signal *(14)*. The high resemblance of the recovered and original signals allows existing data analysis algorithms to continue function without any modification.

The contribution of this work is twofold. First, we have developed a CS-based approach for CV data capturing so that less information is stored and transmitted. Our approach can be easily implemented with current CV data capture techniques. We evaluate our approach using 10 million CV data samples from the SPMD program. As a result, we can recover the CV data with the root mean squared error (RMSE) of 0.05 by keeping only 20% of the original data (i.e., reducing the storage and transmission cost by 80%). Second, we have built a simulation model for a five-mile two-lane freeway segment to compare our approach with two conventional techniques on travel time estimation. As a result, our approach generates the most accurate travel time estimations in all simulation scenarios. In addition, our approach allows a CV with a smaller on-board unit (OBU) capacity to behave similarly to a CV with a higher OBU capacity. This implies that the OBU cost can be reduced via our technique.

## METHODOLOGY

Consider a signal vector $x \in R^N$, which can be represented in terms of a set of orthonormal basis $\{\Psi_i\}_{i=1}^N$, $\Psi_i \in R^N$ as $x = \Psi\alpha$, where $\Psi$ is an $N \times N$ matrix called Sparsifying matrix. The signal $x$ is $K$-sparse if $\alpha$, the transformed coefficient vector, has $K$ nonzero entries. Now, suppose $x \in R^N$ is a vector of CV data samples, e.g., speed samples collected at a fixed rate, then we need to conduct the transform $\alpha = \Psi^T x$ so that $\alpha$ has a sparse representation in the domain of $\Psi$.

Transforms that can fulfill this purpose include discrete Fourier transform (DFT), discrete cosine transform (DCT), and Discrete Wavelet Transform (DWT). DCT is a Fourier-based transform similar to DFT, but uses cosine functions and the transformed coefficients are real numbers. DWT is more suitable for piecewise constant signals *(13)*, which is not applicable to fluctuating speed samples. Therefore, we select DCT to transform a signal *(19)* as

$$\alpha_j = K(j) \sum_{i=1}^{N} x_i \cos \frac{\pi j (i-0.5)}{N}, j = 0, \dots, N-1, \text{ where } K(j) = \frac{1}{\sqrt{N}}, \text{ if } j = 0 \text{ and } K(j) = \sqrt{\frac{2}{N}}, \text{ if } 1 \leq j \leq N-1.$$

Because CS acquires a compressed signal as $y = \Phi x = \Phi \Psi \alpha = \Theta \alpha$, we then need to select a matrix $\Theta$ in order to obtain the sampled vector $y \in R^M$. In addition, $\Theta$ should satisfy the Restricted Isometry Property (RIP) of order $2K$ so that the original vector $x$ can be recovered *(18)*. The previous studies have shown the following theorem:

   **Theorem 1** *(20)*. Suppose a $M \times N$ matrix $\Theta$ is obtained by selecting $M$ rows independently and uniformly at random from the rows of a $N \times N$ unitary matrix $U$. By normalizing the columns to have unit $l_2$ norms, $\Theta$ satisfies the RIP with probability $1 - N^{-O(\delta_{2K}^2)}$ for every $\delta_{2K} \in (0,1)$ provided that $M = \Omega(\mu_U^2 K log^5 N)$, where $\mu_U = \sqrt{N} max_{i,j} |u_{i,j}|$ is the coherence of the unitary matrix $U$.

   Following Theorem 1, we select an $N \times N$ inverse discrete cosine transform (IDCT) matrix $\Psi$ as the unitary matrix $U$, and randomly select its $M$ rows to form the matrix $\Theta$. This allows us to skip the DCT and IDCT transforms and be able to acquire $M$ samples (i.e., $y$) directly from the real observations $x$ as $y = \Theta \alpha = D \Psi \Psi^T x = Dx$, where $\Theta = D\Psi$ and $D$ is a random subset of M rows of a $N \times N$ identity matrix.

   Finally, after determining matrix $\Psi$ and $\Theta$ and assuming a CV is capturing speed samples at a fixed rate, we keep a sample if it is generated from a uniform distribution and less than or equal to the compression ratio $M/N$. When a full signal is needed for certain applications, we can reconstruct it by solving the $l_1$-norm optimization program. We show the applications of our approach in the following two case studies.

**CASE STUDY I: CAPTURE AND RECOVERY OF 10 MILLION BSM SPEED SAMPLES**
Our dataset consists of 16798 continuous trips extracted from the SPMD program. The key parameters of our approach are M and N, which determine the compression ratio thus affecting the recovery accuracy. FIGURE 1 TOP-LEFT shows the RMSEs of the 10 million speed samples different compression ratios. When the compression ratio is 0.1, the RMSE of N=1000 is the lowest. Increasing the compression ratio, the RMSEs become lower and close to each other for different values of N. When the compression ratio is greater than 0.2, the RMSEs are capped at 0.025. When the compression ratio reaches 0.6, all the RMSEs are close to zero.

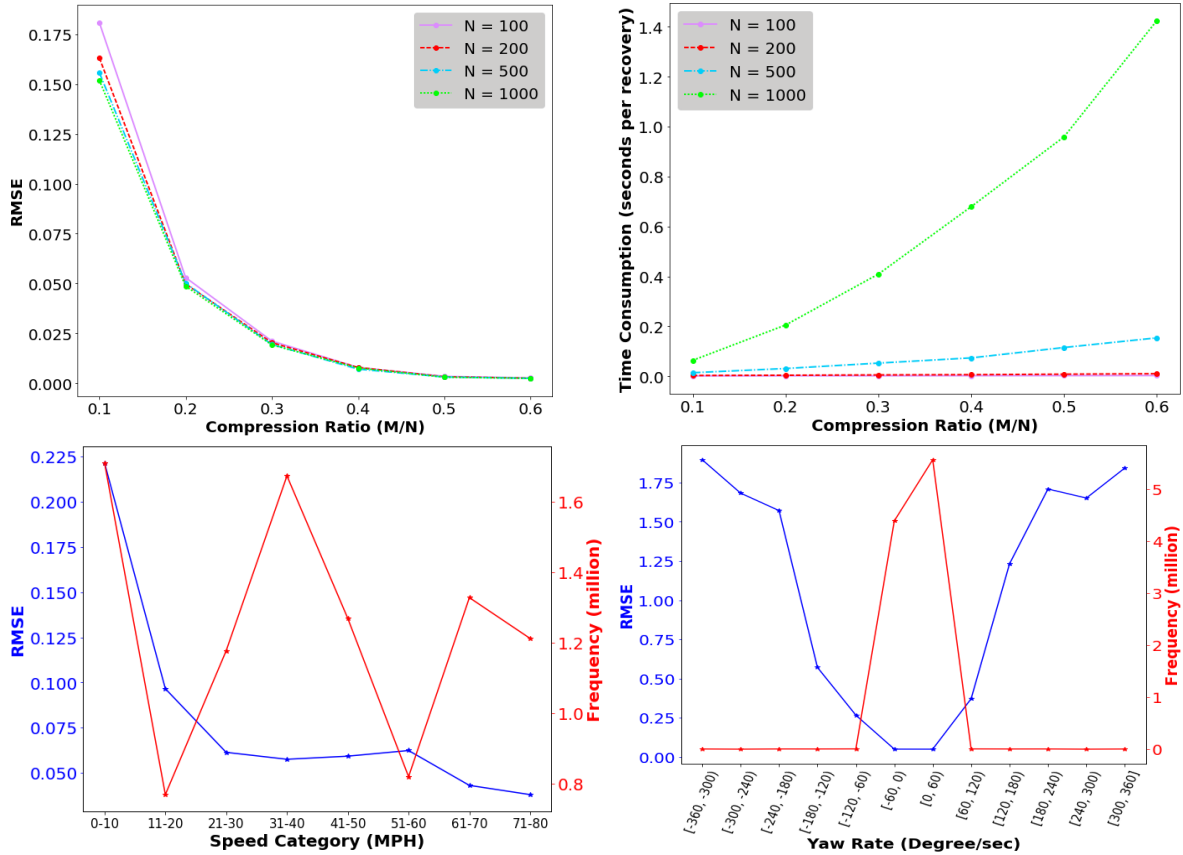

**FIGURE 1 TOP-LEFT: Recovery performance measured by RMSE under various compression ratios. TOP-RIGHT: Recovery performance measured by time per recovery under various compression ratios. BOTTOM-LEFT: Recovery performance by speed category. BOTTOM-RIGHT: Recovery performance by yaw rate category.**

The average time per recovery is also calculated and shown in FIGURE 1 TOP-RIGHT. All the experiments are conducted in Windows 10, i7-6820HK CPU with 64 GB RAM. The time per recovery is close to zero for all the compression ratios when N=100 and N=200. When N=500 and N=1000, the curves of time per recovery have a large increase under larger compression ratios.

Next, we show the recovery performances of our approach related to other BSM variables. We split the original 10 million speed samples into 8 categories by every 10 MPH and calculate the corresponding RMSE and the number of samples within each category. As shown in FIGURE 1 BOTTOM-LEFT, most speed categories have more than 1 million samples except the cases of "11-20" and "51-60". The RMSE curve decreases substantially when the speed category becomes larger. This implies our approach performs better in high speed situations.

Additionally, we evaluate the recovery performance of our approach in 12 yaw rate categories from [-360,360] at 60 degrees interval. FIGURE 1 TOP-RIGHT shows the RMSEs and the number of samples in each category. A negative yaw rate meaning a CV is turning to the left while a positive value indicates turning to the right *(22)*. A large amount of yaw rates falls in [-60, 0) and [0, 60) categories. For other categories, only few thousands of samples exist. The RMSEs are close to zero for the categories of [-60, 0) and [0, 60) and become larger when the yaw rate is higher.

## CASE STUDY II: IMPACT OF OUR APPROACH ON TRAVEL TIME ESTIMATION

Travel time estimation aims to provide the travel time from one point to another in a link for a certain time interval *(23)*. Accurate and reliable travel time estimation plays a critical role in active traffic management *(24,25)*. We conduct our evaluations through the traffic simulator SUMO *(26)*.

For each segment s and each time interval j in the simulation, we can estimate travel times from three data sources: traditional loop detector data, CV data captured at a fixed rate, and CV data via our CS approach. For convenience, these travel times are referred to as $TT_{LP}^{s,j}$, $TT_{CV}^{s,j}$, and $TT_{CS}^{s,j}$. In addition, we denote the obtained the ground truth travel times as $TT_{GR}^{s,j}$. Next, we compute Mean Absolute Percentage Errors (MAPE) of travel time estimations as

$$MAPE_d = \frac{\sum_{s=2}^{N} \sum_{j=4}^{T} \frac{|TT_d^{s,j} - TT_{GR}^{s,j}|}{TT_{GR}^{s,j}}}{(N-1)*(T-3)},$$ where $d = LP, CV, CS$ indicates the three data sources for travel

time estimations. In the following we select 4 scenarios by varying OBU capacity, arrival rate, compression ratio and data capture rate and set the CV market penetration rate to 60%.

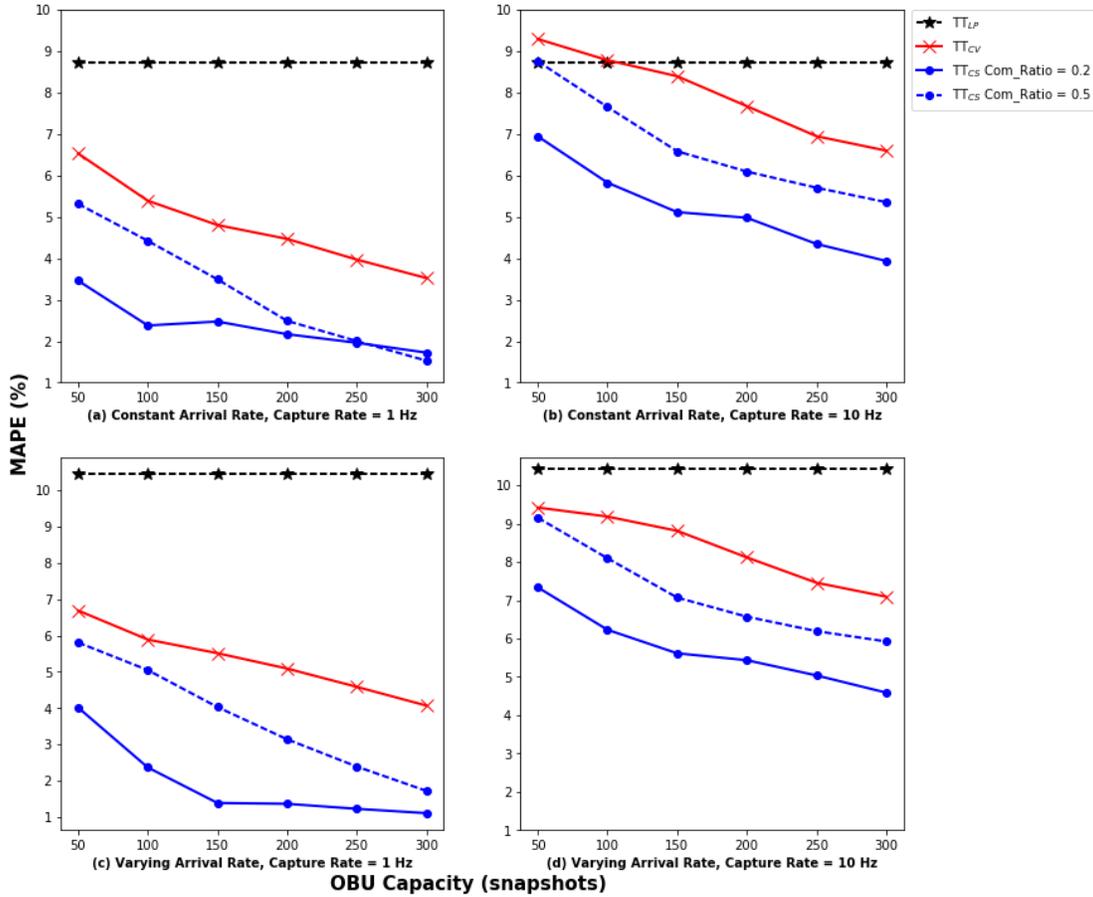

**FIGURE 2 Travel time estimation performances by OBU capacity, arrival rate, compression ratio and data capture rate (CV MPR = 60%).**

First, for most scenarios, $MAPE_{CV}$ and $MAPE_{CS}$ are lower than $MAPE_{LP}$, except a few cases when the data capture rate is 10 Hz and the OBU capacity is 50 or 100 snapshots as shown in FIGURE 2(b). This is mainly because only limited road information can be stored in a CV at a high data capture rate using a small OBU capacity. Second, the $MAPE_{CV}$ and $MAPE_{CS}$ curves are decreasing while the OBU capacity is increasing in all subplots of FIGURE 2. Using the same simulation setting, $MAPE_{CS}$ is always lower than $MAPE_{CV}$. Once again, this is due to the broader spatial-temporal coverage and accurate recovery of the CV data based on the CS approach. The simulation results verify that trading a little accuracy for broader spatial-temporal coverage is beneficial for travel time estimations based on CV data. Third, comparing FIGURE 2(a) to FIGURE 2(b), when the data capture rate changes from 10 Hz to 1 Hz, the MAPE performances of $TT_{CV}$ and $TT_{CS}$ are improved. This indicates a higher CV data capture rate (e.g., 10 Hz) may not be necessary for travel time estimation. Therefore, the hardware cost of OBUs can be saved via our technique.

## CONCLUSION AND FUTURE WORK

We propose a Compressive Sensing (CS) based approach for CV data collection and recovery. Our technique allows CVs to compress the data in real-time and can accurately and efficiently recover the original data. We have demonstrated the effectiveness of our approach using two case studies.

There are several future research directions. First, we would like to test the impact of our approach on other CV applications such as information propagation through real-time V2V communications. The CS approach can also be applied for multi-modal data capture of autonomous vehicle (AV), which has additional sensors such as cameras and LIDAR. Second, the popular network simulator NS-3 will be implemented to build a more realistic traffic simulation model. Last, we would like to consider more factors such as the transmission loss and delay for evaluations.